\documentstyle[aps]{revtex}
\draft
\title{Generalized thermodynamic potentials for
mesoscopic conductors in the presence of transport}
\author{Thomas Christen}
\address{ ABB Corporate Research, \\ 
CH-5405 Baden-D\"attwil, Switzerland}
\begin{document}
\maketitle
\begin{abstract}
It is shown that the nonequilibrium steady-state of a phase-coherent
conductor can be described by a generalized thermodynamic potential
based on the concept of partial densities of states. This is
possible due to the fact that dissipation takes place only in the
contacts for the emitted carriers. Long-range Coulomb interaction
is included, and charge conservation and gauge invariance are
satisfied. The theory is illustrated for a mesoscopic capacitor
with leakage. 
\end{abstract}
\vspace{1cm}

The density of states (DOS) is the basic quantity 
from which equilibrium thermodynamic properties
of a system are derived. Once the DOS is known, 
one can construct an appropriate thermodynamic
potential which has to be minimized in order to 
find the equilibrium state. Due to a production of entropy
in driven systems, on the other hand, nonequilibrium 
states can in general not be obtained by minimizing a
thermodynamic potential. 
In this work I show, however, that generalized 
thermodynamic potentials still exist in mesoscopic
conductors \cite{DATTA} in the presence of transport,
provided that
dissipation takes place only in the reservoirs \cite{BUD}. 
In order to treat the nonequilibrium case,
the concept of the DOS is generalized by introducing
{\it partial densities of states}  \cite{BJPC}
which contain the information from which reservoir the
particles are injected. According to B\"uttiker \cite{BJPC}, 
the specific partial DOS used below are called {\it injectivities}.
The injectivities play an important role in the theory of
time dependent and nonlinear electrical transport in
mesoscopic conductors
\cite{BJPC,BTP,CB}. One can show that the injectivities 
are related to the dwell-time of the particles in the conductor
\cite{BCL,GCB}. The injectivities appear naturally
within the scattering approach to conduction and can be expressed 
in terms of a scattering matrix \cite{BJPC}. Equilibrium thermodynamics
in terms of scattering matrices has been formulated in a different 
context by Dashen, Ma, and Burnstein \cite{DMB}.
Avishai and Band \cite{AVI} discussed the relation between the 
scattering matrix and the total DOS of a
one-dimensional system.\\ \indent
This paper is organized as follows.
First, the definition of the injectivities is recalled for a 
system of non-interacting particles. Secondly, the generalized
thermodynamic potential including the Coulomb interaction is 
constructed for a phase-coherent multi-terminal conductor.
The charge distribution is then determined by minimization of this
potential. Finally, the formalism is applied to
a symmetric capacitor with tunneling between the capacitor 
plates and it is shown that the result is in accordance with
Ref. \cite{CB}.\\ \indent
Consider a phase-coherent conductor connected via contacts 
to reservoirs $\alpha = 1,...,N$ of non-interacting electrons.
Such a conductor is characterized 
by a unitary single-particle scattering matrix
which can be decomposed into submatrices $s_{\alpha\beta}$. The 
indices $\alpha$ and $\beta$ label contacts, and the dimensions of
the submatrices are equal to the
numbers of channels in the associated contacts. The first and 
the second index of $s_{\alpha\beta}$
correspond to out-going and to incoming particles, respectively. 
The scattering matrix is a function of
the energy $E$ of the scattered particle, and is a functional 
of the single-particle potential
$eU(x)$, where $e$ is the electron charge and where $U(x)$ is 
the electric potential. Denoting the trace of a matrix by `Tr',
one can write the transmission probability of a carrier from
contact $\beta $ to contact $\alpha $ 
as $T_{\alpha \beta} = {\rm Tr} (s_{\alpha \beta}^{\dagger} 
s_{\alpha \beta})$. The reflection probability at contact
$\alpha $ is $R _{\alpha} ={\rm Tr} (s_{\alpha \alpha}^{\dagger}
s_{\alpha \alpha})$. At thermodynamic equilibrium,
the response of the particle density to a variation of the Fermi energy
is characterized by the local DOS which can be expressed in 
terms of the scattering matrix elements and
its functional derivatives with respect to the potential \cite{BJPC,GCB}
\begin{equation}
\frac{dn(x)}{dE} = -\frac{1}{4\pi i}\sum _{\alpha \beta} 
{\rm Tr} \left( s_{\alpha \beta}^{\dagger}
\frac{\delta s_{\alpha \beta}}{e\delta U(x)} - 
\frac{\delta s_{\alpha \beta}^{\dagger}}{e\delta U(x)} 
s_{\alpha \beta}\right)\;\; .
\label{eq1}
\end{equation}
Equation (\ref{eq1}) can be interpreted as the sum of {\it injectivities } 
\cite{BJPC}
\begin{equation}
\frac{dn(x,\beta)}{dE} =-\frac{1}{4\pi i}\sum _{\alpha} 
{\rm Tr}\left( s_{\alpha \beta}^{\dagger}
\frac{\delta s_{\alpha \beta}}{e\delta U(x)} - 
\frac{\delta s_{\alpha \beta}^{\dagger}}{e\delta U(x)} 
s_{\alpha \beta}\right)
\label{eq2}
\end{equation}
which are the partial DOS associated with particles injected at
contact $\beta $. In a nonequilibrium situation
where the electrochemical potential $\delta \mu _{\beta}$ in a 
single reservoir is changed, the response of the particle density is given 
by $\delta n(x)=(dn(x, \beta)/dE)\delta \mu _{\beta}$.
A decomposition of the local DOS into injectivities leads to 
the following picture. The total sample is decomposed into
$N$ subsystems. Subsystem $\beta $ consists of all those scattering
states which are associated with particles injected at contact
$\beta $. The local DOS of subsystem $\beta $ is given by the injectivity
$dn(x, \beta)/dE$. In a nonequilibrium steady-state, these 
states are filled as if subsystem $\beta $ were in equilibrium with 
reservoir $\beta $. Dissipation takes place in the reservoirs only for
the out-going particles which are thermalized to the Fermi distribution
of this reservoir. As long as scattering in the conductor
is elastic, dissipation
does not affect the states inside the sample.\\ \indent
For simplicity, I assume zero temperature and consider a discretized
version of a mesoscopic conductor. Importantly, all nearby
conductors and gates are included in this model.
The whole system consists thus of regions  $\Omega _{k}$ ($k=1,...,M$) 
with electrostatic potentials $U_{k}$, charges $q_{k}$, and 
injectivities $D_{k \alpha}$. For later convenience, the 
injectivities are written in the form
\begin{equation}
D_{k \alpha} = e^{2} \int _{\Omega _{k}} d^{3}x\: 
\frac{dn(x,\alpha)}{dE} \;\; ,
\label{eq3}
\end{equation}
which have the dimension of a capacitance. The local
DOS of a single region $\Omega _{k}$ and the total DOS of the sample are
$D_{k}=\sum _{\alpha}D_{k \alpha}$ and $D=\sum _{k \alpha}D_{k \alpha}$,
respectively. If $eV_{\alpha}^{(eq)}$ denotes the equilibrium
electrochemical potential of contact $\alpha $,
the part of the charge on conductor $k $ which is injected 
from contact $\alpha $ for a variation
$\Delta V_{\alpha }=V_{\alpha}-V_{\alpha}^{(eq)}$ of the voltage becomes 
\begin{equation}
q_{k \alpha} = \int _{V_{\alpha}^{(eq)}} ^{V_{\alpha}} 
dV\: D_{k \alpha}(V) \;\; .
\label{eq4}
\end{equation}
The total charge in $\Omega _{k}$ is $q_{k}=\sum _{\alpha}q_{k\alpha}$.
The energy $E_{D}$ of the noninteracting system is given by the sum 
of the energies
over all single-particle states
\begin{equation}
E_{D} = \sum _{k \alpha}\int _{V_{\alpha }^{(eq)}} ^{V_{\alpha }} 
dV\: V D_{k\alpha }(V) \;\;,
\label{eq5}
\end{equation}
where the energy scale is defined such that the equilibrium 
system has zero energy, $E_{D}^{(eq)}=0$. After an expansion of Eqs.
(\ref{eq4}) and (\ref{eq5}) to second order in
$\Delta V_{\alpha }$, one can eliminate the $ \Delta V_{\alpha }$
and express the energy as a function of the partial charges $q_{k \alpha}$
\begin{equation}
E_{D}(\{ q_{k\alpha}\})= \sum _{k, \alpha} \left( 
V_{\alpha }^{(eq)} q_{k\alpha } +
\frac{1}{2}D_{k\alpha }^{-1} q_{k \alpha}^{2} \right) \;\;.
\label{eq6}
\end{equation}
In order to include the long-range Coulomb interaction, it 
is convenient to introduce 
a geometric capacitance matrix $C_{kl}$ for the regions $\Omega _{k}$,
which is determined by the Poisson equation. An arbitrary charge
distribution $\{ q_{l}\}$ induces electrostatic potential 
shifts $U_{k}$ given by 
\begin{equation}
U_{k} = \sum _{l=1}^{N}C^{-1}_{kl}\: q_{l} + U_{0}\;\; .
\label{eq7}
\end{equation}
Note that a global voltage shift $U_{0}$ is always a solution of 
the Poisson equation.
The Coulomb energy of the charge distribution is thus 
\begin{equation}
E_{C}(\{ q_{k\alpha}\})= \frac{1}{2} 
\sum _{k\alpha l\beta} q_{k \alpha} C_{kl}^{-1}q_{l \beta} 
+U_{0}\sum _{k\alpha } q_{k \alpha}\;\; .
\label{eq8}
\end{equation}
At zero temperature, the free energy is
equal to the total energy of the closed system and is
given by the sum of the kinetic
energy $E_{D}$ and the Coulomb energy $E_{C}$
\begin{equation}
E (\{ q_{k\alpha}\}) = \sum _{k \alpha} (U_{0}+V_{\alpha 
}^{(eq)})q_{k \alpha } +
\frac{1}{2} \sum _{k\alpha l\beta } q_{k\alpha }\tilde 
C_{k\alpha l\beta }^{-1}q_{l\beta } \;\; ,
\label{eq9}
\end{equation}
where the following matrix is introduced 
\begin{equation}
\tilde C_{k\alpha l\beta }^{-1} = D_{k\alpha }^{-1} 
\delta _{k l}\delta _{\alpha \beta}+ C_{kl}^{-1} \;\; .
\label{eq10}
\end{equation}
Since the open system is appropriately described by the
grand-canonical ensemble, the potential which must be
minimized is
\begin{equation}
E (\{ q_{k\alpha}\}) -\sum _{k \alpha} 
V_{\alpha } q_{k \alpha } \equiv {\rm Min.} \;\; .
\label{eq12}
\end{equation}
Variation with respect to the charges $q_{k \alpha}$ 
yields finally a set of $M\times N$ equations
\begin{equation}
 \sum _{l\beta } \tilde C_{k\alpha l\beta }^{-1} q_{l\beta }  
= \Delta V_{\alpha}-U_{0} \;\; ,
\label{eq13}
\end{equation}
where $\Delta V_{\alpha}=V_{\alpha}-V_{\alpha}^{(eq)}$ 
corresponds to the (electrochemical) voltage shift
in contact $\alpha$. The still free global shift $U_{0}$ of 
the electric potential is
determined by the additional condition of charge conservation,
\begin{equation}
\sum _{k \alpha} q_{k\alpha }=0 \;\;.
\label{eq11}
\end{equation}
Thus, $U_{0}$ can be interpreted as a Lagrange parameter 
associated with the condition (\ref{eq11}).
Charge conservation is a general property of a complete set of
conductors connected to electron reservoirs, 
since electric fields are fully screened in the reservoirs \cite{BJPC}.
In order to solve Eq. (\ref{eq13}) for the $q_{k \alpha}$,
the matrix $\tilde C_{k\alpha l\beta }$ is introduced as the inverse
matrix of $\tilde C_{k\alpha l\beta }^{-1}$. Note that this 
quadratic matrix acts on $M\times N$-dimensional
vectors $q_{k\alpha}$ (see also the example
below). Combination of Eqs. (\ref{eq13}) and (\ref{eq11}) yields
\begin{equation}
 U_{0} =\frac{ \sum _{k\alpha l\beta }\tilde 
C_{k\alpha l\beta } \Delta V _{\beta}}
 {\sum _{k\alpha l\beta} \tilde C_{k\alpha l\beta } }\;\; .
\label{eq14}
\end{equation}
Once the partial charges $q_{k \alpha}$ are known, one can calculate
the electrochemical capacitance which relates the static 
charge distribution $ q_{k}=\sum _{\alpha}q_{k\alpha}$
to the voltage shifts $\Delta V_{\alpha}$ in the contacts,
\begin{equation}
 C_{\mu ,l \alpha }= \frac{\partial q_{l}}{\partial \Delta V_{\alpha}}
 = \frac{\sum _{k\beta m\gamma n\delta }
 ( \tilde C_{l\beta k\alpha}\tilde C_{m\gamma 
n \delta} - \tilde C_{l\beta n \delta }\tilde C_{m\gamma k\alpha})}
 {\sum _{m\gamma n\delta} \tilde C_{m\gamma n\delta} }  \;\; .
\label{eq15}
\end{equation}
The nonequilibrium electric potential follows from the
charge distribution with the help of Eq. (\ref{eq7}).\\ \indent
An important consequence of charge conservation is that the 
electrochemical capacitance matrix satisfies
$\sum _{l} C_{\mu ,l \alpha}=0$ \cite{BJPC}. 
Additionally, the result is
gauge invariant, $\sum _{\alpha } C_{\mu ,l \alpha}=0$
\cite{BJPC}, i.e. it does not depend on a global voltage shift. The
global voltage shift is absorbed by the constant $U_{0}$ and 
does not change the total energy of the system since the total
charge vanishes.\\ \indent 

As a simple example, consider a symmetric two-terminal capacitor 
with tunneling between the capacitor plates (leakage).
Assume a single one-dimensional open channel with a transmission
probability $T=1-R$. Equivalently, this system describes a symmetric 
quantum point contact with a single open channel, or a one-dimensional
conductor containing a symmetric impurity. This system has already
been discussed in Ref. \cite{CB} in the context of time-dependent
transport. I show that the thermodynamic treatment 
introduced in this work is in accordance with the results of
Ref. \cite{CB}.  Within a semiclassical approximation, the
injectivity from reservoir
$2$ to plate $1$ is proportional to one half of the DOS, 
$D_{1}$ ($=D_{2}$), of plate $1$ and to the transmission probability, $T$. 
The factor one half occurs since 
only particles with a velocity in direction to contact $1$
contribute to this partial DOS. Due to symmetry, it holds 
$D_{12}=D_{21}=D_{1}T/2$ which implies $D_{11}=D_{22}$ $=D_{1}-D_{12}$
$=D_{1}(1-T/2)$ \cite{CB,GCB}. 
The diagonal elements $A\equiv C_{11}^{-1}=C_{22}^{-1}$ and
the off-diagonal elements $B\equiv C_{12}^{-1}=C_{21}^{-1} $ of the
inverse geometrical capacitance matrix are given by 
$A=C_{11}/(C_{11}^{2}-C_{12}^{2})$ and
$B=-C_{12}/(C_{11}^{2}-C_{12}^{2})$. 
The effective (charge conserving) geometric capacitance $C_{0}$
between the two capacitor plates is given by $C_{0}=(C_{11}-C_{12})/2$.
In order to represent the matrix $\tilde C _{k\alpha l \beta}$ 
in a simple way, it is convenient
to abbreviate $K_{1}=D_{11}^{-1}+A$ and  $K_{2}=D_{12}^{-1}+A$. 
The equations (\ref{eq13}) read then
\begin{eqnarray}
\left( \begin{array}{cccc}
K_{1}  & A   & B  & B \\
B & B  & K_{2}   & A  \\
A  & K _{2}  & B  & B \\
B & B  & A   & K_{1}
\end{array} \right) \; 
\left(
\begin{array}{cccc}
q_{11} \\
q_{12} \\
q_{21}  \\
q_{22}
\end{array}
\right) 
=
\left(
\begin{array}{cccc}
 \Delta V_{1}-U_{0} \\
 \Delta V_{1}-U_{0} \\
 \Delta V_{2}-U_{0}  \\
 \Delta V_{2}-U_{0}
\end{array}
\right) 
\;\;\;\;.
\label{ex1}
\end{eqnarray}
By taking charge conservation (\ref{eq11}) into account,
these equations can be solved for the $q_{k \alpha}$. From
the charge $q_{1}=- q_{2}=q_{11}+q_{12}$ on the capacitor
plates one finds the electrochemical capacitance
\begin{equation}
 C_{\mu } = \frac{\partial q_{1}}{\partial \Delta V_{1}}= 
 -\frac{\partial q_{1}}{\partial \Delta V_{2}} 
=\frac{R}{C_{0}^{-1}+4D^{-1}} \;\; .
\label{ex2}
\end{equation}
This result states that, in the case of vanishing transmission,
the electrochemical capacitance is the geometrical 
capacitance in series with quantum corrections given by the DOS
of the plates \cite{BTP}. In the macroscopic limit where the
DOS diverges ($D\to \infty$), one recovers the pure
geometric capacitance, $C_{\mu} = C_{0}$. 
On the other hand, for increasing
transmission the capacitance decreases and vanishes for 
$T=1$. Clearly, without reflection
there is no charge dipole at all. A more extensive discussion 
is provided by Refs. \cite{CB,BCS}. It should be
clear that {\it transport} 
properties such as, e.g., the current cannot be calculated directly
from such a generalized thermodynamic potential which contains
only the information on the charge (or potential) distribution.\\ \indent
In conclusion, I have shown that it is possible to 
describe the stationary nonequilibrium charge distribution of a
mesoscopic phase-coherent conductor with a 
generalized thermodynamic potential. The theory is based on the concept of
partial densities of states (injectivities). The result 
includes the long-range Coulomb-interaction and
is charge conserving and gauge invariant. To illustrate the theory,
I used the simple approximation of a descretized version of the
mesoscopic sample. However, in principle the generalization to a
field theory, e.g., in the framework of a density functional theory,
is straightforward. The generalization to
finite temperatures and to other thermodynamic questions (e.g.,
mechanical properties of the system), remains an interesting future task
but does not require new concepts.
Furthermore, I considered only the deviation from the equilibrium
state described by equal electrochemical potentials. But this is not a
restriction since the equilibrium reference state 
can always be constructed from a thermodynamic potential. 
Although transport properties cannot be extracted directly 
from the generalized potential, it must be emphazised that the
knowledge of the nonequilibrium state is crucial for the
determination of the currents beyond linear dc-response
\cite{BJPC,BTP,CB,LA,CBN}.   


{\em Acknowledgments -}
I thank M. B\"uttiker for helpful discussions.
This work was mainly supported by the Swiss National
Science Foundation under grant Nr. 43966.
%


\begin{thebibliography}{99}

\bibitem{DATTA}   For a review see, e.g., S. Datta, {\em Electronic
                  Transport in Mesoscopic Systems} 
                  (Cambridge University Press, 1995);
                  C. W. J. Beenakker and H. van Houten, 
                  Solid State Phys. {\bf 44}, 1 (1991).
\bibitem{BUD}     M. B\"uttiker, Phys. Rev. B {\bf 33 }, 3020 (1986).
\bibitem{BJPC}    M. B\"{u}ttiker, J. Phys.: Condens. 
                  Matter {\bf 5}, 9361 (1993).
\bibitem{BTP}     M. B\"{u}ttiker, H. Thomas, and 
                  A. Pr\^etre, Phys. Lett. A {\bf 180}, 364 (1993); 
                  Z. Phys. B {\bf 94}, 133 (1994).
\bibitem{CB}      T. Christen and M. B\"{u}ttiker,  
                  Phys. Rev. Lett. {\bf 77}, 143 (1996).    
\bibitem{BCL}     M. B\"uttiker and T. Christen, to appear in
                  {\em Quantum Transport in Semiconductor 
                  Submicron Structures},
                  edited by B. Kramer, (Dordrecht, Kluwer, 1996).
\bibitem{GCB}     V. Gasparian, T. Christen, and M. B\"uttiker, 
                  Phys. Rev. A. (1996).
\bibitem{DMB}     R. Dashen, S.-K. Ma, and H.-J. Bernstein, 
                  Phys. Rev. {\bf 187}, 345 (1969).
\bibitem{AVI}     Y. Avishai and Y. B. Band, Phys. Rev. 
                  B {\bf 32}, 2674 (1985).
\bibitem{LA}      R. Landauer, in {\em Nonlinearity in Condensed Matter},
                  edited by A. R. Bishop et al. 
                 (Springer, Berlin, 1987) p.2.
\bibitem{BCS}     M. B\"uttiker and T. Christen, appears in `Dynamic and 
nonlinear
                  transport in mesoscopic structures', edited by E. Sch\"oll
                  (Chapman and Hall, London).
\bibitem{CBN}     T. Christen and M. B\"{u}ttiker,  Europhys. Lett. (1996).    
\end{thebibliography}
\end{document}